\documentclass{article}

\usepackage{arxiv}

\usepackage[utf8]{inputenc} 
\usepackage[T1]{fontenc}    
\usepackage{hyperref}       
\usepackage{url}            
\usepackage{booktabs}       
\usepackage{amsfonts}       
\usepackage{nicefrac}       
\usepackage{microtype}      
\usepackage{lipsum}
\usepackage{graphicx}
\graphicspath{ {./images/} }

\usepackage{amsmath,amssymb}
\usepackage{bm}
\usepackage{xcolor}
\newcommand{\rr}{{\bm{r}}}
\newcommand{\rrA}{{\bm{r}}_{\rm A}}
\newcommand{\GG}{{\bf{G}}}
\newcommand{\RR}{{\bf{R}}}
\newcommand{\md}{{\rm{d}}}
\newcommand{\me}{{\rm{e}}}
\newcommand{\mi}{{\rm{i}}}
\newcommand{\tr}{{\rm{tr}}}

\newcommand{\jf}[1]{{#1}}

\title{An atom passing through a hole in a dielectric membrane: Impact of dispersion forces on mask-based matter-wave lithography\thanks{This is the version of the article before peer review or editing, as submitted by an author to 
Journal of Physics B: Atomic, Molecular and Optical Physics. IOP Publishing Ltd is not responsible for any errors or omissions in this version of the manuscript or any version derived from it. The Version of Record is available online at https://doi.org/10.1088/1361-6455/ac4b41}}

\author{
 Johannes Fiedler \\
  Department of Physics and Technology\\ University of Bergen\\ All\'egaten 55\\ 5007 Bergen, Norway \\
  \texttt{johannes.fiedler@uib.no} \\
   \And
 Bodil Holst \\
Department of Physics and Technology\\ University of Bergen\\ All\'egaten 55\\ 5007 Bergen, Norway \\
}

\begin{document}
\maketitle
\begin{abstract}
Fast, large area patterning of arbitrary structures down to the nanometre scale is of great interest for a range of applications including the semiconductor industry, quantum electronics, nanophotonics and others. It was recently proposed that nanometre-resolution mask lithography can be realised by sending metastable helium atoms through a binary holography mask consisting of a pattern of holes. However, these first calculations were done using a simple scalar wave approach, which did not consider the dispersion force interaction between the atoms and the mask material. To access the true potential of the idea, it is necessary to access how this interaction affects the atoms. Here we present a theoretical study of the dispersion force interaction between an atom and a dielectric membrane with a hole. We look at metastable and ground state helium, using experimentally realistic wavelengths (0.05-1 nm) and membrane thicknesses (5-50 nm). We find that the effective hole radius is reduced by around 1-7 nm for metastable helium and 0.5-3.5 nm for ground-state helium. As expected, the reduction is largest for thick membranes and slow atoms.
\end{abstract}


\section{Introduction}

Patterning of arbitrary structures in resist on the "few-nanometre" scale (resolution and pitch) can at present only be done using electron beam lithography, ion beam lithography or a scanning probe needle~\cite{Martinez2007}. Common for all these techniques is that they write the pattern in a serial manner, pixel for pixel and thus are too slow for large scale industrial applications. This is a crucial issue for a range of novel technologies such as room temperature quantum electronic devices and room temperature magnetic semiconductors; as well as for quantum electronic devices that use nanoscale effects to do classical computing such as resonant tunnelling diodes, single-electron Coulomb blockade transistors~\cite{Kastner1992,Putnam2017,Wolf2001}, and quantum dot transistors~\cite{doi:10.1063/1.121014}. For these devices to operate at room temperature it is typically required that at least two, and preferably all of the device dimensions, are less than 5 nm. For example, magnetic semiconductors possess two ways of controlling conduction; either via charge carriers, as in conventional semiconductors used in microelectronics, or by the control of quantum spin states, which is an essential ingredient of any spintronics application. By reducing the scale of magnetic materials down to 1~nm dots, ferromagnetic semi-conduction at room temperature is possible. This may potentially permit entirely new kinds of energy-efficient computers and electronic devices because they can be used to make electronic circuits that retain their logic state even after the power is switched off~\cite{Ando2006}. Furthermore, industrial production of new spin-transistors and novel non-volatile magnetic memory concepts will be possible~\cite{Datta1990,Chuang2015}.

Today's lithography industry is dominated by photolithography. State of the art is extreme-ultra-violet lithography (EUV) with a wavelength of 13.5~nm, corresponding to an Abbe resolution limit of 6.75~nm. Normally this resolution limit can be pushed in lithography by playing with overexposure, under-development, and multiple exposures. However, the fundamental problem with using such high-energy photons is that the pattern gets blurred due to secondary electron generation in the resist materials. This blurring puts the ultimate resolution limit at around 6~nm~\cite{Fan2016}. On top of that EUV lithography devices are extremely expensive and hence may not be suitable for smaller volume applications.

It was recently proposed that binary holography with metastable atoms can provide a solution to this challenge~\cite{Nesse2017,Nesse2019}. The concept of binary holography with matter waves was demonstrated experimentally more than 20 years ago using metastable neon atoms~\cite{Fujita1996}. In the new approach, it was shown that by placing the substrate close to the mask and using a mask with nanometre-scale holes, nanometre resolution patterning should be possible~\cite{Nesse2019}. However, the calculations in this paper were done using a simple scalar wave approach, not taking into account the dispersion force interaction between the atoms and the mask-material.

Dispersion forces, such as van der Waals forces between neutral particles or Casimir--Polder forces between neutral particles and dielectric surfaces~\cite{Buhmann12a,Brand2015}, are caused by the ground-state fluctuations of the electromagnetic field. They can be understood via an exchange of virtual photons that are generated as a dipole response of the particle due to the vacuum fluctuation of the field surrounding it. These resulting forces are weak for large separations and dramatically increase with decreasing distances due to the $r^{-4}$-power-law close to dielectric objects. It has been shown that in matter-wave diffraction experiments, dispersion forces can have a significant impact on the experimental results to the point of causing decoherence of the matter wave~\cite{Scheel2012,Fiedler2017}.

\begin{figure}
    \centering
    \includegraphics[width=0.3\columnwidth]{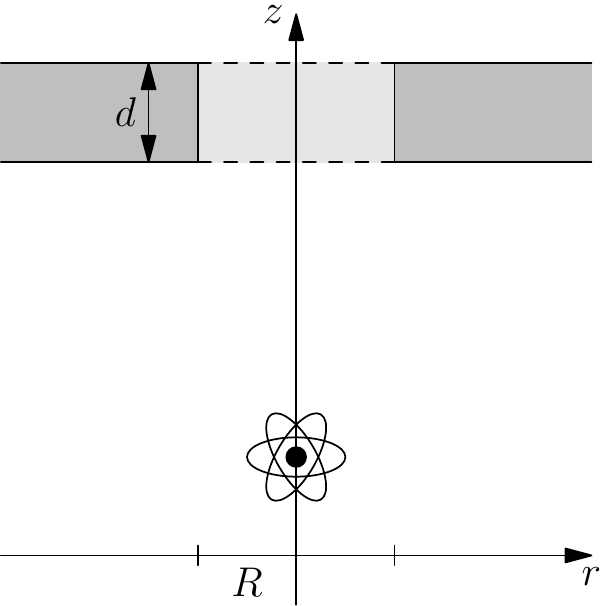}
    \caption{Schematic figure of the considered scenario: An atom with constant velocity ${\bm{v}}=v{\bm{e}}_z =h/(m\lambda_{\rm dB}){\bm{e}}_z$ passes a membrane of thickness $d$ through a hole with radius $R$.}
    \label{fig:schematic}
\end{figure}

\begin{figure}
    \centering
    \includegraphics[width=0.35\columnwidth]{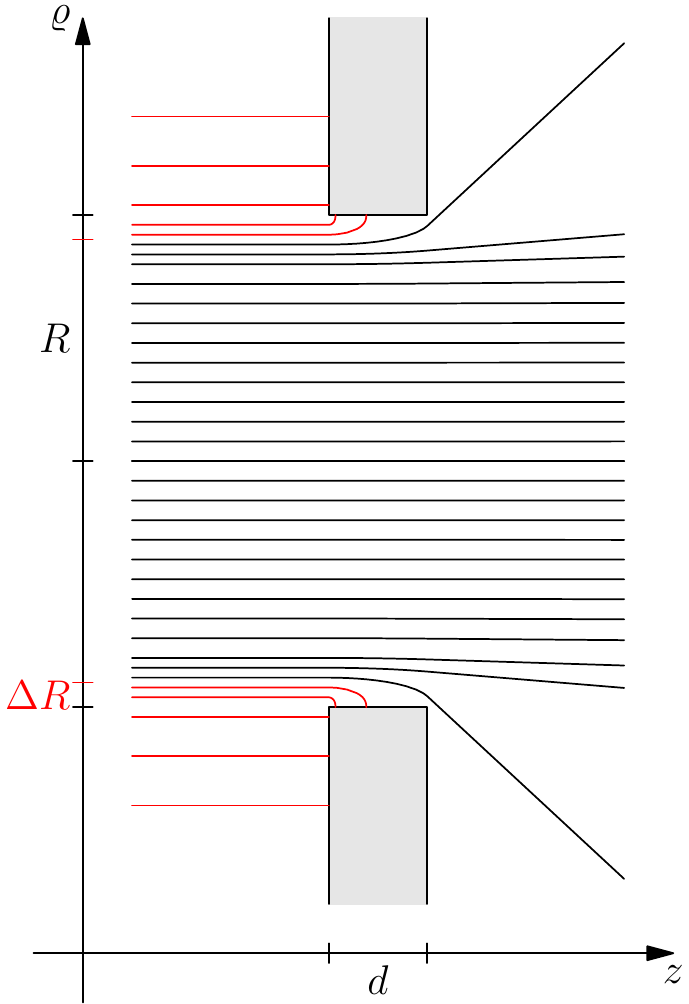}
    \caption{Simulation of the classical trajectories of helium atoms of velocity $v_z=1000\,\rm{m/s}$ (with corresponds to a de-Broglie wavelength $\lambda_{\rm dB} = 0.1\,\rm{nm}$) passing a membrane of thickness $d=50\,\rm{nm}$ through a hole of radius $R=25\,\rm{nm}$. Particles which enter the holes with a distance smaller than or equal $\Delta R$ from the wall of holes will be absorbed.}
    \label{fig:simu_trans}
\end{figure}

In this manuscript, we consider metastable and neutral helium atoms with wavelength $\lambda_{\rm dB}=h/p$ (with the Planck constant $h$, the magnitude of the particle's momentum $p=m\left|{\bm{v}}\right|$ and the mass of the particle $m$) between 0.05 and 1 nm. This is a range that can realistically be made experimentally~\cite{D0CP05833E}. The atoms pass a dielectric membrane through a hole with radius $R$, as depicted in Fig.~\ref{fig:schematic}. The dielectric membrane (in the calculations we use silicon nitride) has a thickness $d$ between 5 and 50 nm, again corresponding to a range that can realistically be made experimentally~\cite{Brand2015}. A matter wave entering the hole can be classified within two different trajectories as depicted in Fig.~\ref{fig:simu_trans}: trajectories which are absorbed by the membrane (red lines) and trajectories that pass the hole. The main purpose of this paper is to determine  the reduction $\Delta R$ of the effective hole radius, which is defined as the radius within which atoms can pass with an accumulated phase shift, $\varphi(\varrho)$. The impact of the Casimir--Polder interaction for a matter wave passing a hole can be described by a complex-valued transmission function~\cite{Juffmann2012,Nimmrichter2008} 
\begin{eqnarray}
t(\varrho) = \Theta[\varrho -\left(R-\Delta R)\right]\me^{\mi\varphi(\varrho)}\,,\label{eq:transmission}
\end{eqnarray}
with the Heaviside function $\Theta$. 

The structure of the paper is as follows: in section 2 we introduce the Casimir--Polder interaction potential for a membrane and determine an effective model describing the potential inside the hole, in section 3 we address the reduction of the hole radius $\Delta R$ via classical motions of particles, where we distinguish between the classical analogue of the eikonal approximation an the full-potential calculation, and in section 4 we model the phase shift $\varphi(\varrho)$ imprinted on a matter wave bypassing a hole due to the dispersion interaction. In section 5, we illustrate the impact of the model on matter-wave diffraction on a single hole.

\section{Dispersion interaction between particle and membrane}

We consider a neutral particle  with polarisability $\alpha$ passing a dielectric membrane with permittivity $\varepsilon$ through a hole with radius $R$. The interaction between the particle and the membrane is determined by the Casimir--Polder potential~\cite{Buhmann12a}
\begin{eqnarray}
U_{\rm CP}(\rrA)= \frac{\hbar\mu_0}{2\pi}\int\limits_0^\infty \md \xi\,\xi^2\alpha(\mi\xi)\tr\left[\GG^{(S)}(\rrA,\rrA,\mi\xi)\right]\,,\label{eq:UCP}
\end{eqnarray}
with the reduced Planck constant $\hbar=h/(2\pi)$ and the vacuum permeability $\mu_0$. In equation~(\ref{eq:UCP}), the scattering Green function $\GG^{(S)}$ denotes the remaining part of the Green function, satisfying the vector Helmholtz equation~\cite{Jackson98}
\begin{equation}
    \left[\nabla\times\nabla\times -\frac{\omega^2}{c^2}\varepsilon(\omega)\right]\GG(\rr,\rr',\omega) = \mathcal{I}\delta(\rr-\rr')\,,\label{eq:Helmholtz}
\end{equation}
with the three-dimensional unit matrix $\mathcal{I}$, by subtracting the bulk (here free space) Green function $\GG^{(0)}$, $\GG(\rr,\rr',\omega) = \GG^{(S)}(\rr,\rr',\omega)+\GG^{(0)}(\rr,\rr',\omega)$. Thus, the Casimir--Polder potential~(\ref{eq:UCP}) can be interpreted as an exchange of virtual photons that are created due to the ground-state fluctuations of the electromagnetic field at the position of the atom $\rrA$ and is back scattered at the dielectric interfaces which are expressed by the spatially dependent dielectric function $\varepsilon(\rr,\omega)$ in the Helmholtz equation~(\ref{eq:Helmholtz}).

A simple example is the Casimir--Polder potential of an atom in front of an infinite half-space, which is described by the spatial permittivity
\begin{equation}
\varepsilon(\rr,\omega) =\left\{  \begin{array}{r@{\quad}cr} 
\varepsilon(\omega) &\mathrm{for}& z<0 \\
    1 &\mathrm{for}& z>0
    \end{array}\right.\,.   
\end{equation}
In this case, the scattering Green function in the coincidence limit ($\rr'\mapsto\rr$) reads~\cite{Buhmann12a,Tomas1995}
\begin{equation}
    \GG^{(S)}_{\rm pl}(\rr,\rr,\mi\xi)=\frac{1}{8\pi^2}\int\frac{\md^2 k^\parallel}{\kappa^\perp}\sum_{\sigma={\rm s,p}}{\bm{e}}_{\sigma+}{\bm{e}}_{\sigma-}\me^{-2\kappa^\perp z}\,, \label{eq:GGpl}
\end{equation}
with the polarisation unit vectors ${\bm{e}}_{\sigma\pm}$ for s- and p-polarised waves, and the imaginary part of the perpendicular wave vector (parallel to the surface normal) 
$\kappa^\perp = \sqrt{\xi^2/c^2+{k^\parallel}^2}$. 

In the non-retarded limit, where the atom-interface distance is smaller than the relevant transition wavelength $z_{\rm A} \ll c/\omega_{\rm max}$ where $\omega_{\rm max}$ is the frequency of the highest relevant atomic transition, \jf{ which is in the order of several tenths of nanometres for Helium. Consequently, retardation effects can be neglected.} The integral in equation~(\ref{eq:GGpl}) can be carried out leading the famous $z_{\rm A}^{-3}$-law 
\begin{equation}
    U_{\rm CP}(z_{\rm A}) = -\frac{C_3}{z_{\rm A}^3} \,,\label{eq:UCPP}
    \end{equation}
    with
    \begin{equation}
    C_3 = \frac{\hbar}{16\pi^2\varepsilon_0}\int\limits_0^\infty \md\xi\alpha(\mi\xi)\frac{\varepsilon(\mi\xi)-1}{\varepsilon(\mi\xi)+1}\,.\label{eq:C3}
\end{equation}

However, this result does not completely map the geometric problem. It is only valid if the particle is close to the interface either in front of the solid plane or near the dielectric surface within the hole. To improve the Casimir--Polder potential with respect to the investigated geometry, the Born-series expansion~\cite{Buhmann12b} can be used, \jf{which is a series expansion of the scattering Green function with respect to the electric susceptibility $\chi=\varepsilon-1$. Its first order} approximates the interaction potential by pairwise interactions \jf{and thus can be applied to electromagnetically weak responding materials such as semiconductors}. The resulting Green function reads~\cite{Bender2014} 
\begin{eqnarray}
\GG^{(S)}_{\rm BS}(\rr,\rr,\mi\xi) = -\frac{\xi^2}{c^2}\frac{\chi(\mi\xi)}{1+\chi(\mi\xi)/3}\int\md^3s \RR(\rr,{\bm{s}},\mi\xi)\cdot\RR({\bm{s}},\rr,\mi\xi)\,,\label{eq:Green_BS}
\end{eqnarray}
where the integral must be taken over the volume of the membrane, with the electric susceptibility $\chi=\varepsilon-1$ and the regular part of the free-space Green function
\begin{eqnarray}
\RR(\rr,\rr',\omega) = \frac{\omega}{4\pi c}\left[f\left(\frac{c}{\omega\varrho}\right)\mathcal{I}-g\left(\frac{c}{\omega\varrho}\right)\frac{\boldsymbol{\varrho}\boldsymbol{\varrho}}{\varrho^2}\right]\me^{\mi \varrho\omega/c}\,,
\end{eqnarray}
with the functions $f(x) = x+\mi x^2-x^3$ and $g(x)=x+3\mi x^2-3x^3$ and the relative coordinate $\bm{\varrho}=\rr-\rr'$. Thus, the trace of the product of both regular parts reads
\begin{eqnarray}
\tr\left[\RR(\rr,{\bm{s}},\mi\xi)\cdot\RR({\bm{s}},\rr,\mi\xi)\right] = \frac{1}{8\pi^2\xi^4\varrho^6 }\me^{-2\varrho\xi/c} \left[\xi^4\varrho^4+2c\xi^3\varrho^3+5c^2\xi^2\varrho^2+6c^3\xi\varrho+3c^4\right]\,,
\end{eqnarray}
which reduces in the non-retarded limit, which is valid for particles closed to the interface, to
\begin{eqnarray}
\tr\left[\RR(\rr,{\bm{s}},\mi\xi)\cdot\RR({\bm{s}},\rr,\mi\xi)\right] = \frac{3c^4}{8\pi^2\xi^4\varrho^6 }\,.
\end{eqnarray}
By integrating over the half space in the Born-series Green function~(\ref{eq:Green_BS}) one obtains the Casimir--Polder potential in the Hamaker approach~\cite{Bender2014}
\begin{eqnarray}
U_{\rm CP,Ham}(z_{\rm A}) = -\frac{\hbar}{16\pi^2\varepsilon_0 z_{\rm A}^3} \frac{9}{6}\int\limits_0^\infty \md \xi\alpha(\mi\xi)\frac{\varepsilon(\mi\xi)-1}{\varepsilon(\mi\xi)+2}\,,
\end{eqnarray}
that deviates from the exact result~(\ref{eq:UCPP}) with (\ref{eq:C3}) by the factor $9/6$ and the Mie reflection coefficient, $(\varepsilon-1)/(\varepsilon+2)$ instead of the Fresnel reflection coefficient, $(\varepsilon-1)/(\varepsilon+1)$. This discrepancy is caused by the single scattering approximation~(\ref{eq:Green_BS}) and can be compensated by including all multiple scattering orders~\cite{Buhmann12b}. For this compensation, one usually introduces an additional correction factor~\cite{Brand2015}. In the non-retarded limit, which is relevant for the considered situation, we explained above, the spatial and frequency dependence of the interaction potential separates and the correction factor can directly be inserted into the potential giving
\begin{equation}
    U_{\rm CP,app}(\rr) = -\frac{9 C_3}{\pi}\int \frac{\md^3s}{\left|{\bm{s}}-\rr\right|^6}\,,\label{eq:Ham}
\end{equation}
with the planar $C_3$ coefficient~(\ref{eq:C3}). The integral volume in Eq.~(\ref{eq:Ham}) is the volume of the membrane. Thus, we approximate the interaction potential between the atom and the membrane by the summation over all pair-wise van der Waals interactions between the atoms and a volume element of the membrane. This approximation is valid near the interfaces~\cite{Hemmerich2016,Gack2020} and gets worse the further we move away from the object. However, the considered scenario in this manuscript requires the potential extremely near the dielectric membrane - the atom is passing through a small hole.

\section{Deflection of classical particles}
In this section, we consider the classical motion of particles passing a hole in a membrane. The classical trajectories are described by Newton's equations of motion, which cannot be solved analytically for the Casimir--Polder potential~(\ref{eq:UCP}) and thus we apply established numerical methods. Here, we consider ground-state and metastable helium atoms passing a membrane made of silicon nitride. The interaction parameters are given in table~\ref{tab:C3}. 

\begin{table}[htb]
    \centering
    \begin{tabular}{c|c}
    Atom & $C_3\,[\rm{meV \,(nm)}^3]$\\\hline
         He & 0.1~\cite{Grisenti1999}  \\
         He$^\star$ & 4.1~\cite{Bruhl2002} 
    \end{tabular}
    \caption{Experimentally obtained Casimir--Polder coefficients for helium and metastable He in front of a silicone nitride surface.}
    \label{tab:C3}
\end{table}

By considering the transmission of a matter wave through an obstacle, one usually restricts oneself to the transverse motion of the particles which corresponds to the eikonal approximation in the wave picture~\cite{Gack2020,Brand2015}. This assumption means that the velocity component due to the acceleration of the particle in the transverse direction is much smaller than the initial longitudinal component. In the first part of this section, we concentrate on the transversal motion inside the hole and neglect the impact of arrival towards and departure from the membrane. In the second part, we investigate the motion in the full potential.
The motion of a classical particle with mass $m$ in a potential $U$ is determined by Newton's equations of motion~\cite{Landau1976Mechanics}
\begin{equation}
    m\ddot{\rr} = -\nabla U(\rr) \,.
\end{equation}
Applying the rotational symmetry by shifting to cylindrical coordinates ($\varrho$, $\varphi$, $z$), one finds an invariance due to the angular component $\varphi$, reducing the complexity to a two-dimensional problem.
\subsection{Classical motion of particles inside the hole}\label{sec:cltrans}
In this part, we restrict our considerations to the potential inside the hole,
\begin{equation}
    U_{\rm hole} = -\frac{C_3}{(R-\varrho)^3}\Theta\left(z^2-\frac{d^2}{4}\right)\,,
\end{equation}
which only depends on the radial component $\varrho$ by neglecting the travelling towards and departure from the hole. Thus, the equation of motion separates into two equations
\begin{eqnarray}
m\ddot{z} &=& 0 \,,\\
m\ddot{\varrho} &=& -\frac{3C_3}{4(R-\varrho)^4}\,,\label{eq:trans}
\end{eqnarray}
with the initial conditions for perpendicular incidence, $z(0)=-d/2$, $\dot{z}(0)=h/m\lambda_{\rm dB}$, $\varrho(0)=r$ and $\dot{\varrho}=0$. Hence, the perpendicular component can be solved analytically and results in
\begin{equation}
    z(t) = \frac{h}{m\lambda_{\rm dB}} t-\frac{d}{2}\,,
\end{equation}
leading to the time of flight to pass the hole $\tau=dm\lambda_{\rm dB}/h$. Figure~\ref{fig:simu_trans} illustrates the trajectories of ground-state helium atoms passing a hole of diameter $2R=50\rm{nm}$ through a $50\,\rm{nm}$ thick membrane made of silicon nitride with a de-Broglie wavelength $\lambda_{\rm dB}=0.1\,\rm{nm}$ modelled by the transverse potential~(\ref{eq:trans}). It can be observed that the hole radius is reduced by $\Delta R$ due to the trajectories terminating at the surface. 

\begin{figure}
    \centering
    \includegraphics[width=0.55\columnwidth]{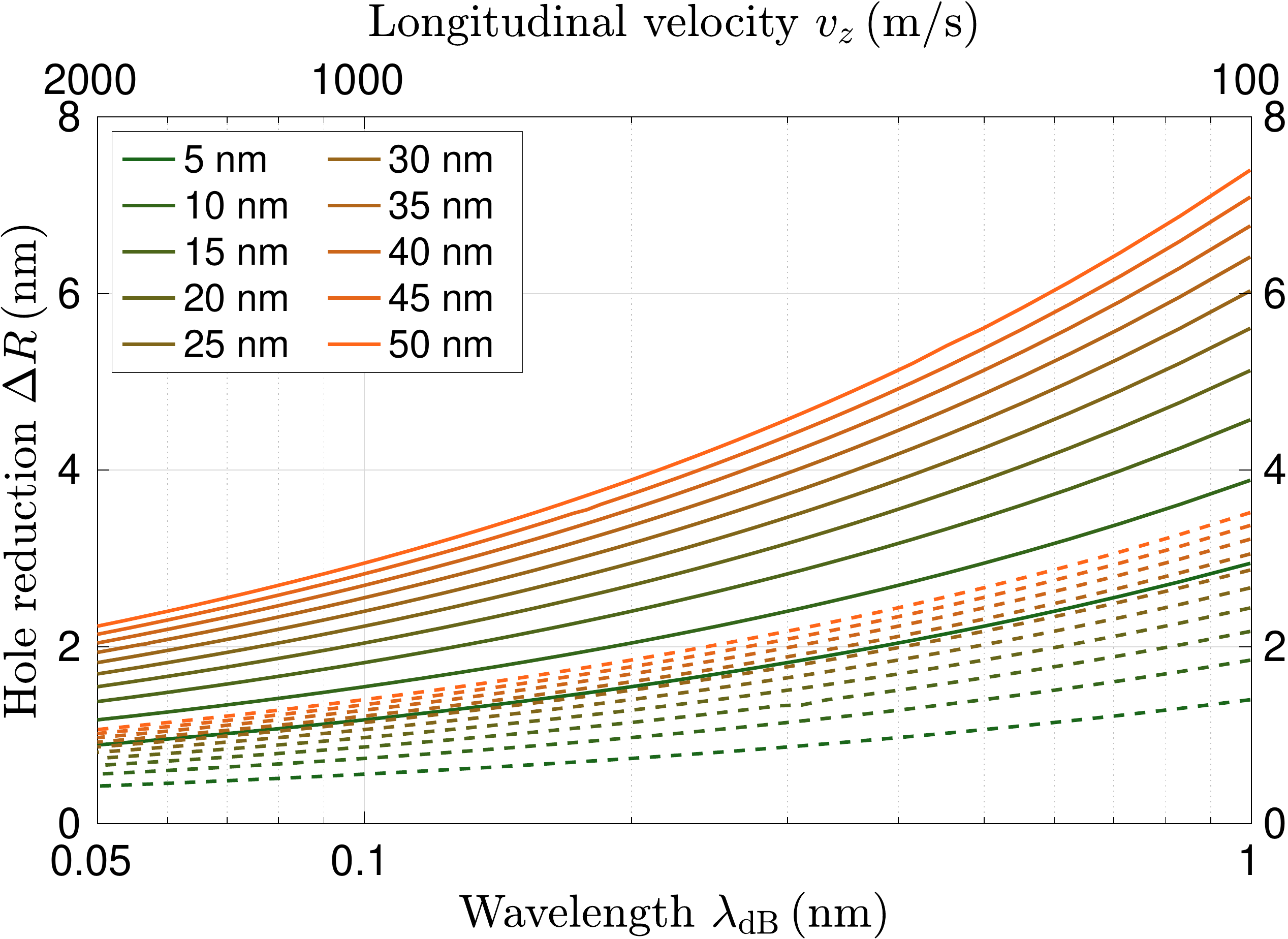}
    \caption{Reduction of the hole's radius $\Delta R$ depending on the de-Broglie wavelength $\lambda_{\rm dB}$ for different membrane thickness from $5$ to $50~\rm{nm}$ for metastable helium atoms (solid lines) and ground-state helium (dashed lines).}
    \label{fig:classical_transversal}
\end{figure}

The reduction of the hole's radius is described by the particle's trajectories with the initial condition $\varrho(0) =R-\Delta R$ and finale point $\varrho(\tau)=R$. Hence, all trajectories entering the hole with $\varrho(0) > R-\Delta R$ will crash into the hole wall, whereas particles entering closer to the hole centre will pass. Due to the form of the attractive force~(\ref{eq:trans}) an analytical solution is not possible. We therefore simulated the classical trajectories and applied a search algorithm to find the critical value $\Delta R$ via a bisection method~\cite{Burden1989}.

The resulting hole reductions are depicted in Fig.~\ref{fig:classical_transversal}. The solid lines correspond to metastable helium and the dashed lines to ground-state helium. It can be observed that the radius is reduced by 1 to 3 nm for metastable helium and by 0.5 to 1.5 nm for ground-state atoms. This difference is obviously due to the difference in the interaction strengths, see table~\ref{tab:C3}. Furthermore, the reduction directly depends on the time of flight, hence faster particles and/or thinner membranes leads to a decrease of the hole reduction.

\subsection{Classical motion of particles through the hole}
\begin{figure}
    \centering
    \includegraphics[width=0.6\columnwidth]{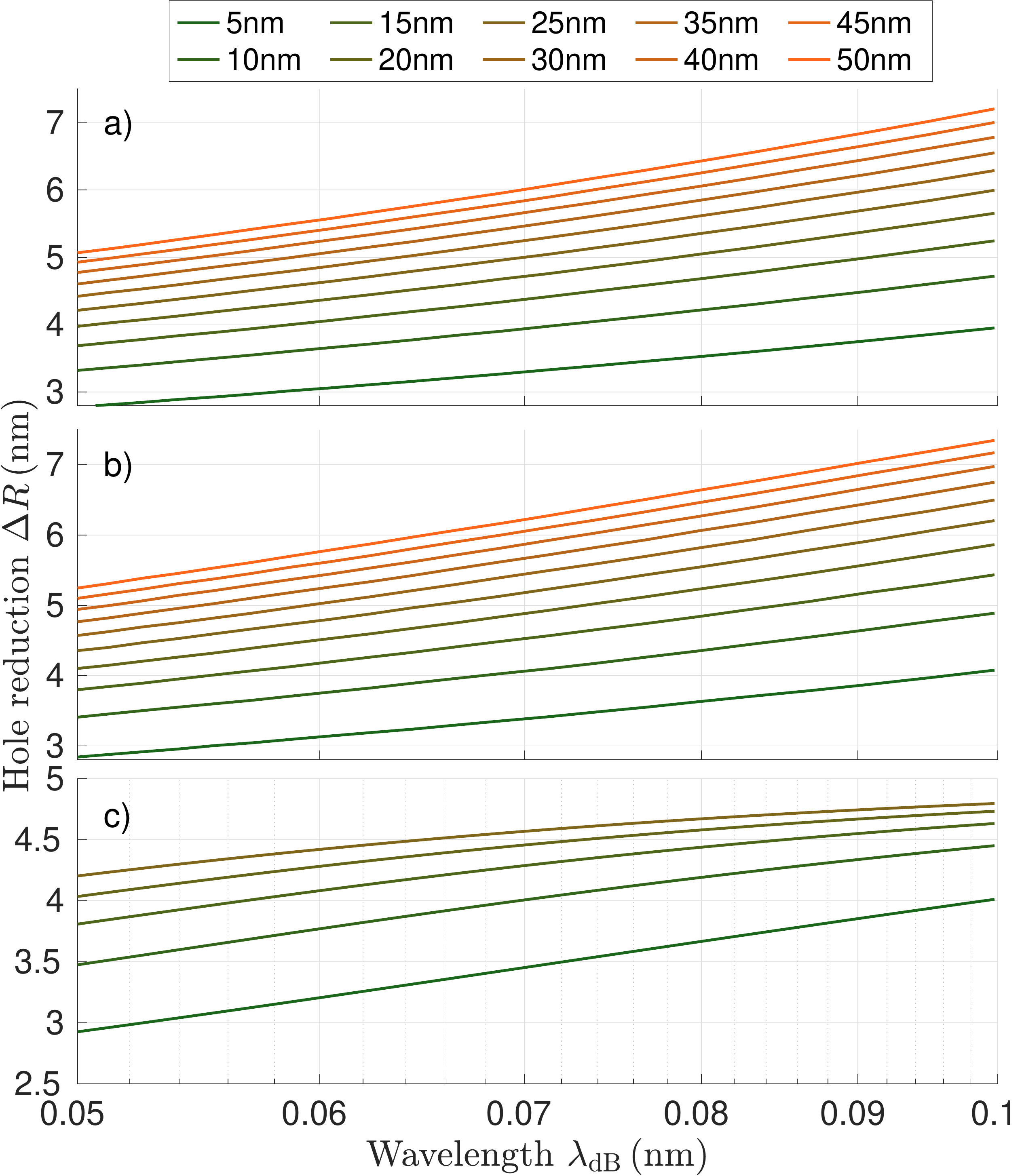}
    \caption{Hole reduction for meta-stable helium atoms passing a silicone nitride membrane with radius 25 nm (a), 10 nm (b) and 5 nm (c) for different membrane thicknesses from 5 to 50 nm (green to red lines).}
    \label{fig:advanced}
\end{figure}

The model described in the previous section~\ref{sec:cltrans} only considers the transverse motion of the atoms. This means that the wave still propagates as a plane wave through the potential landscape. For the description of interference in matter-wave optics, this is a common approach and is described by the eikonal approximation. However, its validity for the hole reduction is not given in general and hence its explicit consideration is required. In the picture of classical particles, this means that the full interaction potential needs to be analysed. Based on the results of the previous section~\ref{sec:cltrans}, it is sufficient to analyse the validity for the parameters that lead to the strongest interaction, which means metastable atoms (largest Casimir--Polder coefficient, see table~\ref{tab:C3}). The Casimir--Polder potential for the three-dimensional case can be approximated by the Hamaker approach~(\ref{eq:Ham}). Thus, the integral in Eq.~(\ref{eq:Ham}) must be carried out (in cylindrical coordinates) over the volume bounded by $R\le r\le\infty$, $-d/2\le z\le d/2$ and $0\le\varphi\le2\pi$. The absorption of particles is numerically hard to handle due to the divergence of the interaction potential near the interface. In general, one defines a cut-off region around the absorbing object and tests the results for different values, similar to what was done for quantum reflection in Ref.~\cite{QR2017}. In our case, the results depend on the size of the cut-off region because we are interested in the absorption of particles at the interface. For this reason, we followed a different approach to compare the purely transversal simulations, see sec.~\ref{sec:cltrans} with the full interaction potential by defining a cut-off region due to the physical relevance by the maximal deflection angle $\beta$ induced by the interaction 
\begin{equation}
    \tan\beta = \frac{v_\perp}{v_z} = \frac{\lambda_{\rm dB}}{\lambda_{\rm ind}}\,,
\end{equation}
as the ratio between the finale perpendicular ($v_\perp$) and longitudinal velocity ($v_z$) which corresponds to the ratio between the de-Broglie wavelength and the wavelength induced by the interaction $\lambda_{\rm ind}$. This approach is motivated by the measurability of the transmitted atoms at a detector/resist-coated substrate, which typically does not cover the entire spherical angle. This approach can be applied to both scenarios (the purely transversal potential, see Sec.~\ref{sec:cltrans}, and the full potential, this section) so that a direct comparison is possible. We simulated the particle's trajectories for different values for the maximal deflection angle $\beta$ in the range of $1$ to $100\,\rm{mrad}$ (which corresponds to transverse induced wavelengths between 0.1 and 10\% of the initial wavelength) and compared the resulting hole reductions. Only a small number of particles are affected by a stronger deflection. Figure~\ref{fig:advanced} illustrates the resulting hole reduction for a maximal deflection angle of $\beta=10\,\rm{mrad}$. All tested angles showed similar behaviour. It can be observed that for holes with radii larger than 10 nm, the reduction stays the same. For smaller radii, the effective interaction strength increases due to the stronger curvature of the hole wall. The simple approximation in sec.~\ref{sec:cltrans} threats the inner hole surface as a plane located at the shortest distance and perpendicularly orientated. This approximation obviously fails for smaller radii and yields a further reduction of about 60\% compared to the eikonal approximation. Interestingly, the reduction stays constant for different deflection angles. 

\section{Phase shift of matter waves}
Due to the duality of waves and particle, the diffraction at a dielectric interface can be described via classical waves, which can be derived via Kirchhoff's diffraction formula~\cite{BornWolf:1999:Book} that determines the propagation of a classical wave with amplitude $a_0$
\begin{eqnarray}
\psi_{\rm{cl}} = \frac{a_0 k_0}{2\pi \mi}\int \md A\, t_{\rm h} \frac{\me^{\mi k_0\left(r_{\rm S,h}+r_{\rm h,D}\right)}}{r_{\rm S,h}r_{\rm h,D}} \frac{\cos\vartheta+\cos\vartheta'}{2}\,, \label{eq:Kirchhoff}
\end{eqnarray}
through an interface with the transmission function $t_{\rm h}$ (subscript h for hole) via the propagation of spherical waves from the source $\rr_{\rm S}$ (subscript S for source) to the hole $\rr_{\rm h}$ along the distance $r_{\rm S,h}=\left|\rr_{\rm S} - \rr_{\rm h}\right|$, continued by the propagation to the detector $\rr_{\rm D}$ (subscript D for detector) along the distance $r_{\rm h,D} =\left|\rr_{\rm h}-\rr_{\rm D}\right|$. The wave vector is related via the de-Broglie wavelength $k_0 = 2\pi/\lambda_{\rm dB} = p/\hbar$ with the particle's momentum $p$. The geometric correction angles $\vartheta$ and $\vartheta'$ are the angles between the aperture's normal ${\bm{n}}$ and the wave's propagation directions, $\rr_{\rm S,h}$ and $\rr_{\rm h,D}$, respectively. 

By considering plane waves passing the hole, the propagation lengths are dominated by the distances between the source and the interface $r_{\rm S,h}\approx L_1$ and between the interface and the detector $r_{\rm h,D} \approx L_2$. This approach simplifies Kirchhoff's diffraction formula~(\ref{eq:Kirchhoff}) to the Fourier transform of the transmission function and is known as Fraunhofer approximation
\begin{eqnarray}
\psi_{\rm{cl}} = \frac{a_0 k_0}{2\pi\mi}\frac{\me^{\mi k_0 \left(L_1 + L_2 + p^2/(2L_2)\right)}}{L_1L_2}\int\md A\, t_{\rm p} \me^{\mi \frac{k_0}{L_2}{\bm{p}}\cdot{\bm{s}}}\,, \label{eq:Fraunhofer}
\end{eqnarray}
where ${\bm{p}}$ describes the in-plane coordinates of the detector ${\bm{r}}_{\rm D} = ({\bm{p}},0)$ and ${\bm{s}}$ of the interface ${\bm{r}}_{\rm p} = ({\bm{s}},0)$. Due to the rotational symmetry of the hole, the use of cylindrical coordinates is recommended, and the angular integral can be carried out leading to
\begin{eqnarray}
\psi_{\rm{cl}}(p) = -\mi a_0 k_0\frac{\me^{\mi k_0 \left(L_1 + L_2 + p^2/(2L_2)\right)}}{L_1L_2}\int\limits_0^R\md \varrho \,\varrho t_{\rm p}(\varrho) J_0\left( \frac{k_0}{L_2}p\varrho\right)\,.\label{eq:Hankel}
\end{eqnarray}
Due to the interaction between a particle and the interface, the transmission function needs to be modified by the interaction potential~(\ref{eq:Ham}) which is typically considered by a phase shift of the wave bypassing the interface, in analogy to the transmission function~(\ref{eq:transmission}),
\begin{equation}
    \varphi(\varrho) = - \frac{1}{\hbar} \int U\left[{\bm{r}}(t)\right]\md t\,,\label{eq:phase}
\end{equation}
which describes the phase accumulated along the particle's trajectories ${\bm{r}}(t)$ to the initial condition $r(0)=\varrho$ due to the interaction potential. We assume normal incidence, thus not include the geometric phase which is, for instance, caused by tilting the incoming plane wave. 

In the eikonal approximation, the particle's trajectories are considered as straight lines, along which the variance of the longitudinal velocity can be neglected $z=v_z t=ht/(m\lambda_{\rm dB})$. This approach yields to the substitution
\begin{equation}
    \varphi(\varrho) \approx - \frac{m\lambda_{\rm dB}}{2\pi\hbar^2} \int U(\varrho,z) \md z\,.
\end{equation}
Due to the rapid spatial decrease of the interaction potential~(\ref{eq:Ham}) in front and after the membrane, the integration range can be extended along the complete $z$-axis ($-\infty<z<\infty$). In this approach, the spatial integrals can be carried out and the phase shift reads
\begin{eqnarray}
\varphi(\varrho) =-\frac{m \lambda_{\rm dB}}{2\pi\hbar^2}\frac{3 C_3 d}{4 R^3 \left(\lambda-1\right)^3\left(\lambda+1\right)^2}\left[ \left(\lambda-1\right)^2 K\left(\frac{2\sqrt{\lambda}}{\lambda+1}\right) -\left(\lambda^2+7\right)E\left(\frac{2\sqrt{\lambda}}{\lambda+1}\right)\right],\label{eq:eiko}
\end{eqnarray}
with $\lambda=\varrho/R$ and the elliptic integrals
\begin{equation}
    K(x) = \int\limits_0^{\pi/2}\frac{\md\vartheta}{\sqrt{1-x^2\sin^2\vartheta}}\,,
\end{equation}
and
\begin{equation}
    E(x) = \int\limits_0^{\pi/2}\sqrt{1-x^2\sin^2\vartheta}\,\md\vartheta\,.
\end{equation}

By comparing the approximated phase shift with the one obtained by following the particle's trajectories, it can be observed that for the relevant particles, which are outside the cut-off region defined in the previous section, the deviation between the phase shifts in eikonal approximation~(\ref{eq:eiko}) and along the particle's trajectories~(\ref{eq:phase}) is in the order of a few per cent for most of the particles. Only a small number of particles passing the hole near the wall will collect a stronger phase shift. However, they will also be absorbed by the surface of the membrane. 

\jf{
The derived model considers perfectly spherical holes in a membrane. However, in practice, a produced pattern of holes will, of cause, contain imperfections, such as non-circular holes; surface roughness; surface vibrations; and electrostatic potentials caused by charge accumulations induced by fabrication and/or grain boundaries. An adaptation of the model to include these imperfections is possible in principle, but enormously extends the parameter space. The first two types of imperfections (non-circularity and surface roughness) are stationary effects and can be characterised by, for instance, the eccentricity $e$ (for ellipsoidal shaped holes) and an effective amplitude $A$ and periods $\omega_z$ (along the main axis of the hole) and $\omega_\varphi$ (along the angular direction). These parameters define a deformed surface area of the hole and, thus, can be included in the interaction potential via boundary conditions of the volume integral~(\ref{eq:Ham}). Furthermore, the defined volume must be projected onto the $\varrho$-$\varphi$-plane to calculate the integral range for the Fraunhofer diffraction integral~(\ref{eq:Fraunhofer}). A similar analysis concerning a wedge angle of grating bars has been performed in Ref.~\cite{Brand2015}. Surface vibrations are a dynamic effect and hence require a time series of different hole shapes. Phonon vibrations caused by the finite temperature of the grating are already included in the $C_3$-coefficient due to their contribution to the dielectric response of the material. Strong temperature differences between the membrane and the helium atom led to a radiative heat transfer~\cite{Schmidt_2018}, which can also lead to the decoherence of matter waves~\cite{Cotter2015,PhysRevA.93.063612}. Electrostatic forces can be considered via an additional potential due to the induced dipole and permanent dipole interaction $U(\rrA)=-\alpha(0) \left|{\bm{E}}(\rrA)\right|^2/2-{\bm{d}}\cdot{\bm{E}}({\bm{\rrA}})$~\cite{Fiedler2017,Bender2014}, where the latter vanishes for ground-state Helium, leading to an additional phase shift bypassing the hole.}
\begin{figure}
    \centering
    \includegraphics[width=0.6\columnwidth]{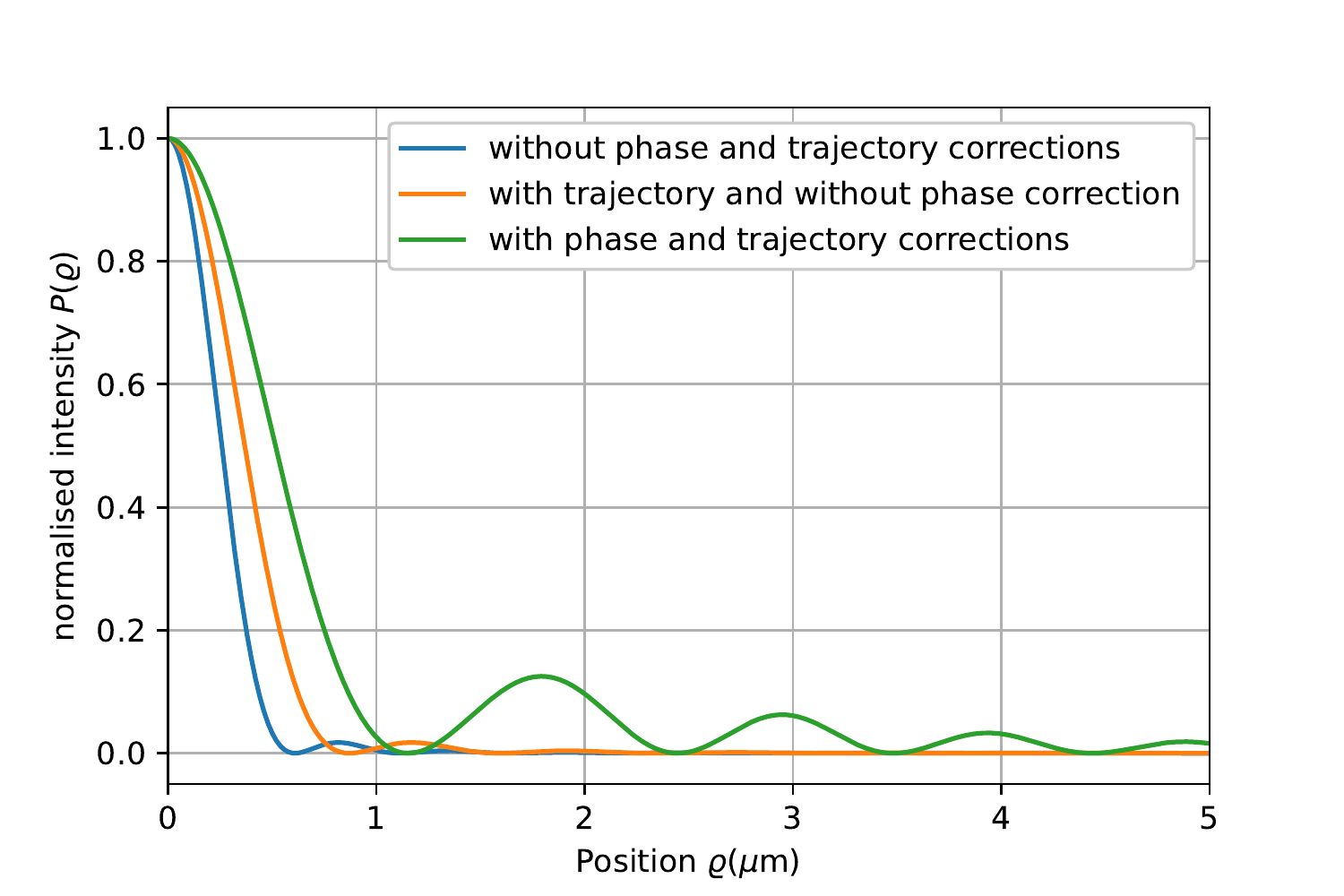}
    \caption{Normalised interference pattern for the diffraction of metastable helium at a hole with radius $R=5\,\rm{nm}$ in a $5\,\rm{nm}$ thick silicone nitride membrane.}
    \label{fig:interference}
\end{figure}
\section{Impact of the dispersion forces on the diffraction at a hole}
To illustrate the impact of the Casimir--Polder interaction on an expected interference pattern, we consider the far-field diffraction of metastable helium at a silicon nitride membrane of thickness $d=5\,\rm{nm}$ and a hole radius $R=5\,\rm{nm}$. We consider a moderate wavelength of the incoming particle $\lambda_{\rm dB}=0.1 \,\rm{nm}$. We set the length between the source and the membrane to be $1\,\rm{m}$ and between the membrane and the screen to $50\,\rm{\mu m}$ to ensure the far-field regime and an incoming plane which allows the application of Eq.~(\ref{eq:Hankel}) to determine the interference pattern. The resulting interference patterns are depicted in Fig.~\ref{fig:interference}. The patterns have been normalised relative to the corresponding zeroth-order intensity. It can be observed that the higher diffraction orders are higher populated and stronger diffracted due to the correction. A further important aspect is the reduction of the total transmission rate by 3.5\%. To distinguish between the two effects: --- hole reduction and phase shift --- we estimated an interference pattern by considering the reduction of the hole radius only and neglecting the phase shift. Remarkably, the transmission rate only decreases to 0.8\% due to the effective hole radius. The remaining intensity loss is caused by the stronger diffraction due to the phase shift.

\section{Conclusion}
This paper shows that dispersion forces lead to a non-negligible trajectory broadening and phase shift when neutral ground-state or metastable helium atoms, passes holes of dimensions applicable to matter-wave mask lithography. The effective hole radius is reduced by around 1 nm for typical experimental conditions (0.05 nm wavelength, 5 nm membrane thickness). We addressed this issue via the classical particle's motion and the propagation of classical waves. These techniques are commonly applied in matter-wave optics. However, future work should include a full, quantum-mechanical consideration of the transmission of particles. This treatment would also address the adsorption of particles. Future theoretical work on mask-based matter-wave lithography should include the dispersion force-induced trajectory broadening and phase-shift described in this paper \jf{as well as typical experimental imperfections, such as non-circular holes; surface roughness; surface vibrations; and electrostatic potentials as discussed in Sec. 4}. 
\section*{Acknowledgements}
This work was funded by the European
Union’s Horizon 2020 research and innovation
programme H2020-FETOPEN-2018-2019-2020-01 under Grant Agreement No. 863127 nanoLace (www.nanolace.eu).
\bibliographystyle{unsrt}  
\bibliography{bibi.bib}  

\end{document}